\documentclass[aps,showpacs,epsfig,twocolumn]{revtex4}
\usepackage{epsfig}

\newcommand{\be}{\begin{equation}}
\newcommand{\ee}{\end{equation}}
\newcommand{\bea}{\begin{eqnarray}}
\newcommand{\eea}{\end{eqnarray}}
\usepackage{color}
\begin{document}

\title{\bf\Large {Ferromagnetic Quantum critical behavior  in three-dimensional Hubbard model with transverse anisotropy}
}

\author{Naoum Karchev }

\affiliation{ Department of Physics, University of Sofia, 1126 Sofia, Bulgaria }

\begin{abstract}
One-band Hubbard model with transverse anisotropy is considered at density of electrons $n=0.4$.
It is shown that when the anisotropy is appropriately chosen, the ground state is ferromagnetic with magnetic order perpendicular to the anisotropy.  The  increasing of the ratio $\frac tU$, where $t$ is the hopping parameter and $U$ is the Coulomb repulsion, decreases the Curie temperature, and the system arrives at the quantum critical point $(T_C=0)$. The result is obtained introducing  Schwinger bosons  and slave Fermions representation of the electron operators. Integrating out the spin-singlet Fermi fields an effective Heisenberg model with ferromagnetic exchange constant is obtained for vectors which identifies the local orientation of the spin of the itinerant electrons. The amplitude of the spin vectors is an effective spin of the itinerant electrons accounting for the fact that some sites, in the ground state, are doubly occupied or empty.
Owing to the anisotropy, the magnon fluctuations drive the system to quantum criticality and when the effective spin is critically small these fluctuations suppress the magnetic order.

\end{abstract}

 \maketitle

Quantum phase transitions (QPT) arise in many-body systems because of competing interactions
that support different ground states. At quantum critical point (QCP)the matter undergoes
a transition from one phase to another at zero temperature.
A nonthermal control parameter, such as pressure, doping or magnetic field, drives the system to QCP.
Quantum phase transitions are a subject of great interest \cite{Hertz,Sachdev99,MVojta,Stockert,Hilbert07}.
At this point, the quantum critical fluctuations give rise to unconventional temperature dependence of
magnetic, thermal, and transport parameters \cite{Hertz,Millis}.

Quantum phase transition can be induced in a wide range of materials. Most prominent experimental realization of
ferromagnetic quantum phase transition is found in magnetic properties of $LiHoF_4$ \cite{Bitko}. At low temperature
the magnetic degrees
of freedom in this material are the spins of the holmium atoms.
They have an easy axis, and below $T_C=1.53K$ the compound is ferromagnet. In \cite{Bitko} the authors measured the
magnetic order as a function of temperature and a magnetic field applied perpendicular to the easy axis.
The increasing of the magnetic field reduces Curie temperature monotonically. When it is larger than some critical field (about $50 kOe$)
the long-range order in the material is destroyed even at zero temperature. The spin flip operators are the
quantum fluctuations which drive the system to the quantum critical point \cite{Bitko,MVojta}. Another examples of field-induced quantum phase transitions are discussed in the review articles \cite{Sachdev99,MVojta,Stockert}.

Pressure driven transformation of $CeRu_2Ge_2$ from ferromagnet into paramagnet, at zero temperature, is studied in \cite{Sullow}. The suppression of magnetic order at the critical pressure $p_{cr}=67kbar$ is accompanied by non-Fermi-liquid behavior.

The magnetic properties of $URh_{1-x}Ru_xGe$ alloys are investigated in \cite{Huy}.
The Curie temperature vanishes linearly with $"x"$ and the ordered moment is suppressed in a continuous way at $x_{cr}=0.38$.
The thermal, transport, and magnetic properties of $URh_{1-x}Ru_xGe$ near the critical concentration are investigated.
The data provide evidence for continuous ferromagnetic quantum phase transition.

The earliest theory of a ferromagnetic quantum phase transition was the Stoner theory\cite{Stoner}. In later paper \cite{Hertz} Hertz derived an effective Ginzburg-Landau-Wilson theory starting from a microscopic theory of itinerant electrons with four-fermion interaction. He concluded that
in the dimensions $d=2$ and $d=3$ the quantum phase transition from paramagnet to ferromagnet in isotropic system is a second-order transition with mean-field critical behavior. In contrast to this it was shown \cite{Belitz99,Belitz03,Chubukov04} that the quantum phase transition from a metallic paramagnet to an itinerant ferromagnet in $d=2$ and $d=3$ is discontinuous. It was proved that this conclusion is true and for anisotropic systems with magnetic order parallel to anisotropy axis \cite{Belitz12}.

In the present paper the three-dimensional Hubbard model with transverse anisotropy is investigated. The experimental observations in $LiHoF_4$  inspire
that the anisotropy could be of grate importance for the existence of a quantum critical behavior in the system. It is shown that when the anisotropy is appropriately chosen, the ground state is ferromagnetic with magnetic order perpendicular to the anisotropy. The  increasing of the ratio $\frac tU$, where $t$ is the hopping parameter and $U$ is the Coulomb repulsion, decreases the Curie temperature. Owing to the anisotropy, the spin flip fluctuations (magnons) drive the system to quantum criticality and when the effective spin, which accounts for the fact that some sites, in the ground state, are doubly occupied or empty, is critically small these fluctuations suppress the magnetic order.
Critically high double occupancy is phenomena of basic relevance to second-order quantum phase transition in itinerant magnets.

We consider a theory with Hamiltonian
\bea \label{QCBFM1}
h = & - & t\sum\limits_{\langle ij \rangle} \left( c_{i\sigma }^ + c_{j\sigma}  + h.c. \right)
+ U \sum\limits_i n_{i\uparrow} n_{i\downarrow} -\mu \sum\limits_i n_i \nonumber  \\
&-& J' \sum\limits_{  \langle  ij  \rangle }[ S^x_iS^x_j+\kappa S^y_iS^y_j+S^z_iS^z_j]
\eea
where $c_{i\sigma }^+$ and $c_{i\sigma }$ ($\sigma=\uparrow,\downarrow$) are creation and annihilation operators for spin-1/2 Fermi operators of itinerant electrons, $n_{i\sigma}=c^+_{i\sigma}c_{i\sigma}$, $n_i=n_{i\uparrow}+n_{\downarrow}$,
$t>0$ is the hopping parameter, $U>0$ is the the Coulomb repulsion,  and $\mu$ is the chemical potential. The exchange constant in the Heisenberg term is ferromagnetic  $J'>0$, $\kappa$ is a parameter of anisotropy, and the spin of the itinerant electron is \be\label{QCBFM2}
S^{\nu}_i=\frac 12 \sum\limits_{\sigma\sigma'}c^+_{i\sigma} \tau^{\nu}_{\sigma\sigma'}c^{\phantom +}_{i\sigma'}\ee
where $\tau^{x},\tau^{y},\tau^{z}$ are the Pauli matrices.
The sums in Eq.(\ref{QCBFM1}) are over all sites of a three-dimensional cubic lattice, and
$\langle i,j\rangle$ denotes the sum over the nearest neighbors.

We represent the Fermi operators, the spin of the itinerant electrons and the density operators $n_{i\sigma}$  in terms of the Schwinger bosons
($\varphi_{i,\sigma}, \varphi_{i,\sigma}^+$) and slave Fermions
($h_i, h_i^+,d_i,d_i^+$). The Bose fields
are doublets $(\sigma=1,2)$ without charge, while Fermions
are spinless with charges 1 ($d_i$) and -1 ($h_i$).
\begin{eqnarray}\label{QCB2} & & c_{i\uparrow} =
h_i^+\varphi _{i1}+ \varphi_{i2}^+ d_i, \qquad c_{i\downarrow} =
h_i^+ \varphi _{i2}- \varphi_{i1}^+ d_i, \nonumber
\\
& & n_i = 1 - h^+_i h_i +  d^+_i d_i,\quad  S^{\nu}_i=\frac 12
\sum\limits_{\sigma\sigma'} \varphi^+_{i\sigma}
{\tau}^{\nu}_{\sigma\sigma'} \varphi_{i\sigma'},\nonumber
\\& &
c_{i\uparrow }^+c_{i\uparrow }c_{i\downarrow }^+c_{i\downarrow}=d_i^+d_i \eea

\be\label{QCB2b}
\varphi_{i1}^+ \varphi_{i1}+ \varphi_{i2}^+ \varphi_{i2}+ d_i^+
d_i+h_i^+ h_i=1  \ee

To solve the constraint (Eq.\ref{QCB2b}),which entangles the Fermi and Bose operators, one makes a change of variables, introducing
Bose doublets $\zeta_{i\sigma}$ ($\zeta^+_{i\sigma}$) \cite{Schmeltzer}
\begin{eqnarray}\label{QCB3}
\varphi_{i\sigma} & = & \zeta_{i\sigma}\left(1-h^+_i h_i-d^+_i
d_i\right)^
{\frac 12},\nonumber \\
\varphi^+_{i\sigma} & = & \zeta^+_{i\sigma}\left(1-h^+_i h_i-d^+_i
d_i\right)^ {\frac 12}.
\end{eqnarray}
Now, only the new Bose fields are constrained
$\zeta^+_{i\sigma}\zeta_{i\sigma}\,=\,1$. In terms of the new fields
the spin vectors of the itinerant electrons have the form
\be
S^{\nu}_{i}=\frac 12 \sum\limits_{\sigma\sigma'} \zeta^+_{i\sigma}
{\tau}^{\nu}_{\sigma\sigma'} \zeta_{i\sigma'} \left[1-h^+_i
h_i-d^+_i d_i\right] \label{QCB4} \ee
When, in the ground state,
the lattice site is empty, the operator identity $h^+_ih_i=1$ is
true. When the lattice site is doubly occupied, $d^+_id_i=1$. Hence,
when the lattice site is empty or doubly occupied the spin on this
site is zero. When the lattice site is neither empty nor doubly
occupied ($h^+_ih_i=d^+_id_i=0$), the spin equals $\,\,{\bf s}_{i}=1/2
{\bf n}_i,\,\,$, where the unit vector
\be\label{QCB44}
n^{\nu}_i=\sum\limits_{\sigma\sigma'} \zeta^+_{i\sigma}
{\tau}^{\nu}_{\sigma\sigma'} \zeta_{i\sigma'}\qquad ({\bf
n}_i^2=1)\ee identifies the local orientation of the spin of the
itinerant electron.

The Hamiltonian Eq.(\ref{QCBFM1}), rewritten in terms of Bose fields Eq.(\ref{QCB3}) and slave Fermions, adopts the form
\bea\label{QCB3a}
h  & = & -t\sum\limits_{\langle ij \rangle} \left[\left ( d^+_j d_i-h^+_j h_i \right) \zeta^+_{i\sigma}\zeta_{j\sigma}\right. \nonumber \\
& + & \left.\left ( d^+_j h^+_i-d^+_i h^+_j\right )\left (\zeta_{i1}\zeta_{j2}-\zeta_{i2}\zeta_{j1}\right ) + h.c. \right]\nonumber \\
& \times & \left(1-h^+_i h_i-d^+_id_i\right)^{\frac 12}\left(1-h^+_j h_j-d^+_jd_j\right)^{\frac 12} \nonumber \\
& + & U \sum\limits_i d^+_id_i -\mu \sum\limits_i \left (1-h^+_ih_i+d^+_id_i\right)\nonumber \\
& - & J'\sum\limits_{  \langle  ij  \rangle }[ n^x_in^x_j+\kappa n^y_in^y_j+n^z_in^z_j] \\
& \times & \left(1-h^+_i h_i-d^+_id_i\right)\left(1-h^+_j h_j-d^+_jd_j\right)\nonumber .\eea

An important advantage of working with Schwinger bosons and slave Fermions
is the fact that Hubbard term is in a diagonal form. The fermion-fermion and fermion-boson interactions are included in the hopping and spin exchange Heisenberg terms. To proceed we approximate the hopping term of the Hamiltonian Eq.(\ref{QCB3a}) setting  $\left(1-h^+_i h_i-d^+_id_i\right)^{\frac 12}\sim 1$ and keep only the quadratic, with respect to Fermions, terms. This means that the averaging in the subspace of the Fermions is performed in one fermion-loop approximation. Further, we represent the resulting Hamiltonian as a sum of two terms
\be\label{QCB4a}
h=h_0 + h_{int}, \ee
where
\bea\label{QCB4b}
h_0 = & - & t\sum\limits_{\langle ij \rangle} \left ( d^+_j d_i-h^+_j h_i + h.c.\right)
 +  U \sum\limits_i d^+_id_i \nonumber \\
& - & \mu \sum\limits_i \left (1-h^+_ih_i+d^+_id_i\right),\eea
is the Hamiltonian of the free $d$ and $h$ fermions, and
\bea\label{QCB4c}
h_{int} = & - &t\sum\limits_{\langle ij \rangle} \left[\left ( d^+_j d_i-h^+_j h_i \right) \left (\zeta^+_{i\sigma}\zeta_{j\sigma}-1\right)\right. \\
& + & \left.\left ( d^+_j h^+_i-d^+_i h^+_j\right )\left (\zeta_{i1}\zeta_{j2}-\zeta_{i2}\zeta_{j1}\right ) + h.c. \right]\nonumber \\
& - & J'\sum\limits_{  \langle  ij  \rangle }[ n^x_in^x_j+\kappa n^y_in^y_j+n^z_in^z_j] \\
& \times & \left(1-h^+_i h_i-d^+_id_i\right)\left(1-h^+_j h_j-d^+_jd_j\right)\nonumber \eea
is the Hamiltonian of boson-fermion interaction.

The ground state of the system, without accounting for the spin fluctuations, is determined by the free-fermion Hamiltonian $h_0$ and is labeled by the density of electrons
\be\label{QCB4d} n=1-<h^+_i h_i>+<d^+_id_i> \ee (see equation (\ref{QCB2})) and the "effective spin" of the electron
\begin{equation}
m=\frac 12 \left(1-<h^+_i h_i>-<d^+_id_i>\right). \label{QCB5}
\end{equation}

At density of electrons $n=0.4$ one calculates the chemical potential $\mu$ as a function of $t/U$ and utilizes
the result to calculate the effective spin of the itinerant electron $"m"$ as a function of the control parameter $t/U$.
The result is depicted in Fig.(\ref{QCBFMfig1})-black triangles-left scale.

Let us introduce the vector,
\begin{equation}
 M^{\nu}_{i}= m \sum\limits_{\sigma\sigma'} \zeta^+_{i\sigma}
{\tau}^{\nu}_{\sigma\sigma'} \zeta_{i\sigma'}\quad {\bf M}_{i}^2=m^2 .
\label{QCB7}
\end{equation}
Then, the spin-vector of itinerant electrons Eq.(\ref{QCB4}) can be written in the
form
\be\label{QCB5a} {\bf S}_{i}=\frac {1}{2m}{\bf M}_{i}\left(1-h^+_i\,h_i\,-\,
d^+_i\,d_i\right),\ee
where the vector  ${\bf M}_i$ identifies the local orientation of the spin of
the itinerant electrons.
The contribution of itinerant electrons to the total magnetization
is  $<{\bf S}^z_{i}>$. Accounting for the definition of $m$ (see
Eq.\ref{QCB5}), one obtains $<{\bf S}^z_{i}>= <{\bf M}^z_{i}>$.

The Hamiltonian is quadratic with
respect to the Fermions $d_i, d^+_i$ and $h_i, h^+_i$, and one can
average in the subspace of these Fermions (to integrate them out in
the path integral approach). As a result, one obtains an effective
model for vectors  ${\bf M}_i$ with Hamiltonian
\bea\label{QCBFM-eff-h1}
 h_{eff} = & - & J'' \sum\limits_{  \langle  ij  \rangle  } {{\bf M}_i
\cdot {\bf M}_j} \\
& - & J'\sum\limits_{  \langle  ij  \rangle }[ M^x_iM^x_j+\kappa M^y_iM^y_j+M^z_iM^z_j]. \nonumber
\eea
The effective exchange constant $J''$ is calculated in the one
fermion-loop approximation, in the limit when the frequency and the wave
vector are small. At zero temperature, one obtains
\bea\label{QCB10} J'' & = &
\frac {t}{6m^2}\frac 1N
\sum\limits_{k}\left(\sum\limits_{\nu=1}^3\cos
k_{\nu}\right)\left[\theta(-\varepsilon^d_k)-\theta(-\varepsilon^h_k)\right]
 \\
& - & \frac {2t^2}{3m^2 U}\frac 1N
\sum\limits_{k}\left(\sum\limits_{\nu=1}^3\sin^2
k_{\nu}\right)\left[1-\theta(-\varepsilon^h_k)-\theta(-\varepsilon^d_k)\right]\nonumber\eea
where $N$ is the number of lattice's sites, $\varepsilon^h_k$ and
$\varepsilon^d_k$  are Fermions' dispersions,
\bea\label{QCB11}
\varepsilon^h_k & = & 2t(\cos k_x+\cos k_y+\cos k_z) +\mu  \\
\varepsilon^d_k & = & -2t(\cos k_x+\cos k_y+\cos k_z)+U -\mu,
\nonumber \eea
and the wave vector $k$ runs over the first Brillouin zone of a cubic lattice.
The $"m"$ dependence of $J''$ is a consequence of the definition of the vector $M_i$ Eq.(\ref{QCB7}),
and $J'' m^2$ doesn't depend on "m".

It is convenient to rewrite the effective hamiltonian in the form
\be\label{QCBFM-eff-h2}
 h_{eff} =  -  J\sum\limits_{  \langle  ij  \rangle }[ M^x_iM^x_j+r M^y_iM^y_j+M^z_iM^z_j],
\ee
with exchange constant $J=J''+J'$ and effective parameter of anisotropy $r=(J''+\kappa J')/(J''+J")$.

The exchange constant $J$, the effective spin $m$ and the effective anisotropy parameter $r$ are functions of the ratio $t/U$. They are
calculated at density of itinerant electrons $n=0.4$, for anisotropy parameter $\kappa=-0.83$ and $J'/t=10$.
The functions are depicted in Fig.(\ref{QCBFMfig1}).
\begin{figure}[h]
\centerline{\psfig{file=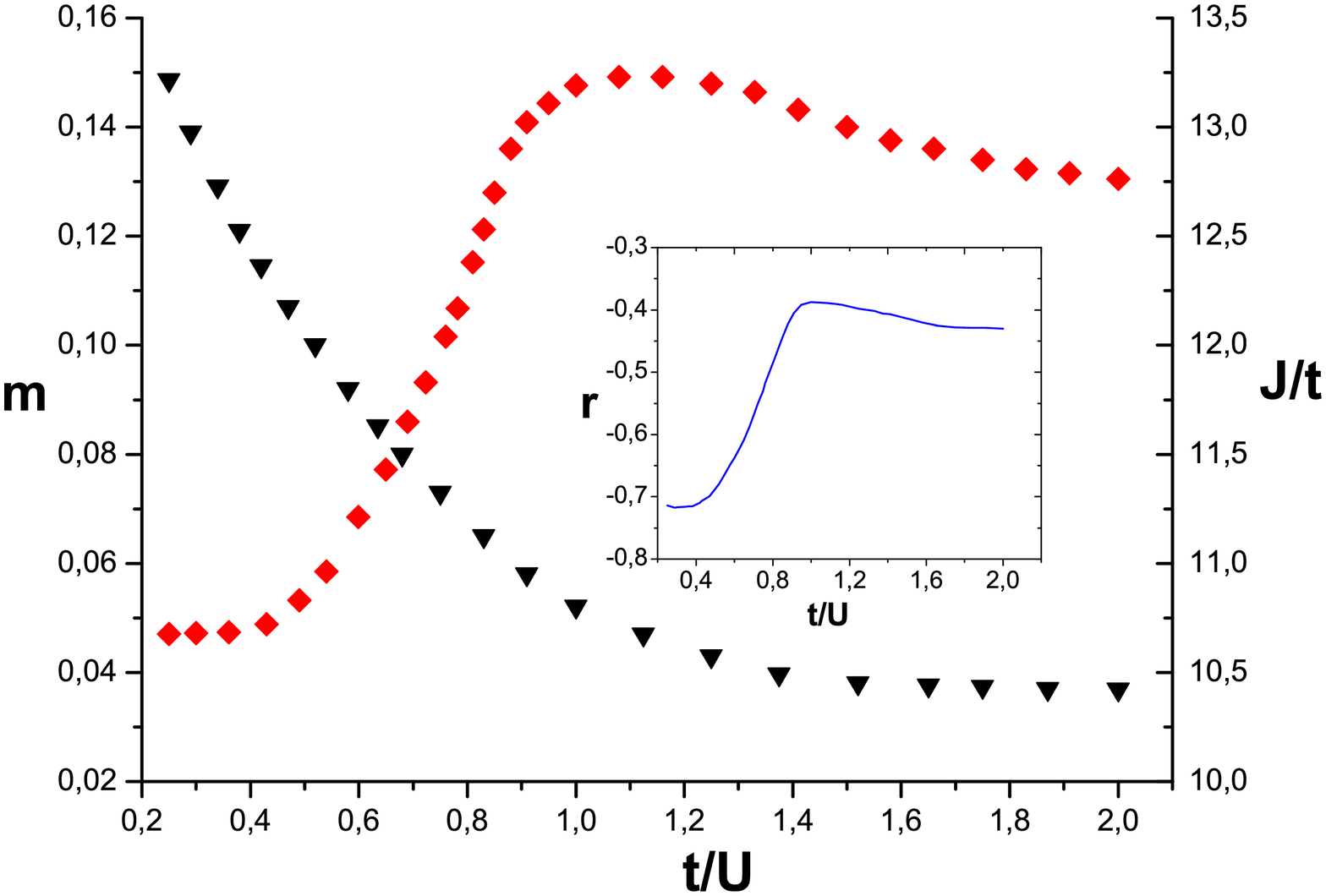,width=9cm,height=7cm}}
\caption{(Colour on-line) The effective spin of the itinerant electrons $\textbf{m}$ as a function of the control parameter $t/U$-black triangles(left scale)
The exchange constant $J/t$ as a function of $t/U$-red rhombuses(right scale). Inset: The effective anisotropy constant $\textbf{r}$ as a function of $t/U$. The functions are calculated at density of itinerant electrons $n=0.4$, for anisotropy parameter $\kappa=-0.83$ and $J'/t=10$.}
\label{QCBFMfig1}
\end{figure}

For chosen parameters the exchange constant is positive $J>0$ and a ferromagnetic order along z axis, perpendicular to the anisotropy, is stable. To study this phase one introduces the Holstein-Primakoff representation of the spin vectors ${\bf M}_j(a^+_j,\,a_j)$
\begin{eqnarray}\label{QCB14}
M_j^+ & = & M_j^x+iM_j^y
 = \sqrt{2m-a_j^+a_j}\, a_j \nonumber \\
M_j^- & = & M_j^x-iM_j^y
 = a_j^+ \sqrt{2m-a_j^+a_j}\nonumber \\
M_j^3 & = & m-a_j^+ a_j ,
\end{eqnarray}
where $a^+_j,\,a_j$ are Bose fields and $m$ is the effective spin of itinerant electrons. In spin-wave approximation, in momentum space representation the effective Hamiltonian
(Eq.\ref{QCBFM-eff-h2}) adopts the form
\bea\label{QCBFM-eff h3}
h_{eff} & = & \sum\limits_{k}\left [\varepsilon_k a_k^+a_k\,-\,\gamma_k \left (a_k^+a_{-k}^+\,+\,a_ka_{-k}\right ) \right ] \nonumber \\
\varepsilon_k & = & Jm\left[6-(1+r)\left (\cos k_x + \cos k_y +\cos k_z \right )\right]\nonumber \\
 \gamma_k & = & Jm\frac {(1-r)}{2} \left (\cos k_x + \cos k_y +\cos k_z \right ). \eea
 To diagonalize the Hamiltonian one introduces new Bose field
$\alpha_k,\,\alpha_k^+$:
$ a_k\,=u_k\,\alpha_k\,+\,v_k\,\alpha^+_{-k}$,
where the coefficients $u_k$ and $v_k$ are real functions of the wave vector.
The transformed Hamiltonian adopts the form \be
\label{QCB29} h_{eff} = \sum\limits_{k}\left
(E_k\,\alpha_k^+\alpha_k\,+\,E^0_k\right),
\ee
with dispersion
\bea\label{QCB30}
E_k & = & 2Jm\sqrt{\left(3-\delta_k \right)\left(3-r\delta_k \right)} \nonumber \\
     \delta_k & = & \cos k_x + \cos k_y +\cos k_z \eea
and vacuum energy
$E^{0}_k\,=\,\frac
12\,\left [E_k\,-\,\varepsilon_k \right]$.

The dispersion Eq.(\ref{QCB30}) shows that the ferromagnetism with magnetic order perpendicular to the anisotropy axis is stable if the effective anisotropy constant satisfies the condition $-1<r<1$. The inset in Fig.(\ref{QCBFMfig1}) shows that all calculated values of the effective anisotropy constant satisfy the above condition.

The dispersion is equal to zero at $\textbf{k}=(0,0,0)$. Therefor, $\alpha_k$-Boson describes  long-range excitation (magnon) in the system.
Near the zero vector the dispersion adopts the form $E_k\propto c_s |\textbf{k}|$
with spin-wave velocity
\be\label{QCBFMcs} c_s=Jm\sqrt{6(1-r)}. \ee
The unusual, for ferromagnetism, dispersion is in consequence of anisotropy. When the isotropy is restored ($r=1$) one obtains the well known ferromagnetic dispersion.

The magnetization is defined by the equation (\ref{QCB14}).
At Curie temperature $T_C$, the magnetization is zero and one calculates the critical temperature solving the equation
\be\label{QCBFM-TC}
2m+1= \frac 1N \sum\limits_{k}\frac {\varepsilon_k }{E_k}\left[1+\frac {2}{e^{\frac {E_k}{T_C}}-1}\right].\ee
When the parameter of anisotropy $r$ is fixed, the critical temperature decreases, decreasing the effective spin $m$. At quantum critical point ($T_C=0$) one obtains a dependence of the critical value of the effective spin $m_{cr}$ on the anisotropy parameter. The relation is given by the equation
 \be\label{QCBFM-mcr}
2m_{cr}+1=\frac {1}{2N} \sum\limits_{k}\frac {6-(1+r)\delta_k}{\sqrt{(3-\delta_k)(3-r\delta_k)}}\ee
with $\delta_k$ from Eq.(\ref{QCB30}), and is depicted in Fig(\ref{m-r}).
\begin{figure}[h]
\centerline{\psfig{file=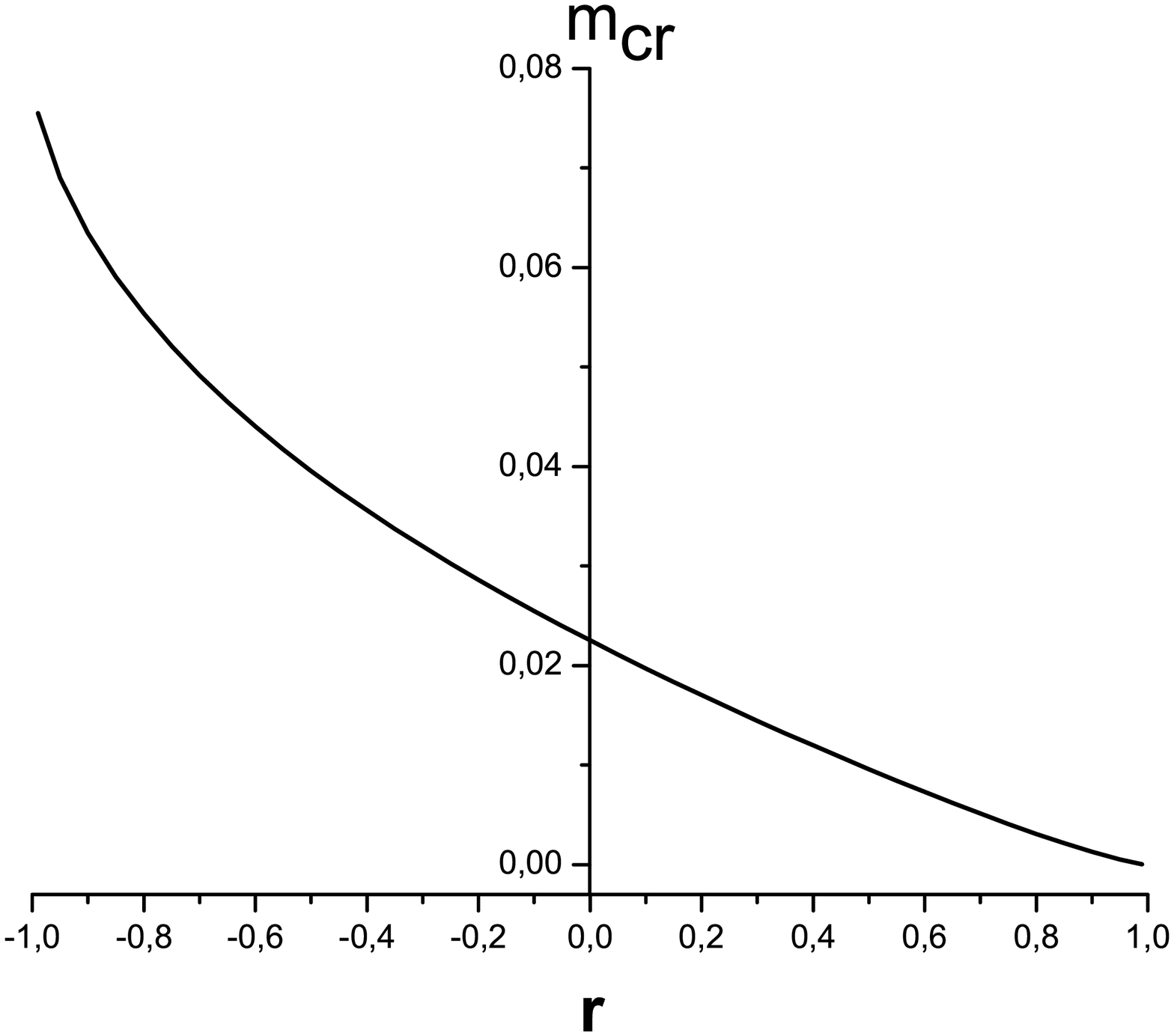,width=8cm,height=6cm}}
\caption{(Colour on-line)\,\,The effective spin $m_{cr}$, at quantum critical point, as a function of the effective parameter of anisotropy.}
\label{m-r}
\end{figure}

The figure (2) demonstrates  very well the nature of the quantum criticality in itinerant ferromagnets with anisotropy. At quantum critical point the effective spin, which is the fermion contribution to the magnetization without accounting for the spin fluctuations, is nonzero. Magnon fluctuations suppress this magnetization to zero. Hence, the magnon (spin flip) fluctuations drive the system to quantum critical point. This is a second order quantum phase transition from ferromagnetic phase with long-range magnon fluctuations to paramagnetic phase with gapped magnons.

This is in contrast to quantum phase transition in isotropic  itinerant ferromagnets ($r=1$). For this systems the quantum phase transition is from ferromagnetic phase with long-range magnon fluctuations to paramagnetic phase without spin-flip fluctuations ($m_{cr}=0$). This explains the non-second order nature of the transition.

Utilizing the dependence of the effective spin $m$ and the exchange constant $J/t$ on the control parameter $t/U$, see Fig. (\ref{QCBFMfig1}), one can obtain the dependence of the dimensionless temperature $T_C/t$ on the ration $t/U$. The phase diagram in the plane of temperature $T_C/t$ and control parameter $t/U$ is depicted in figure (\ref{QCBFMfig3}). At density of electrons $n=0.4$,  anisotropy parameter $\kappa=-0.83$ and $J'/t=10$ the quantum critical value of the ratio is $t/U=2$.

\begin{figure}[h]
\centerline{\psfig{file=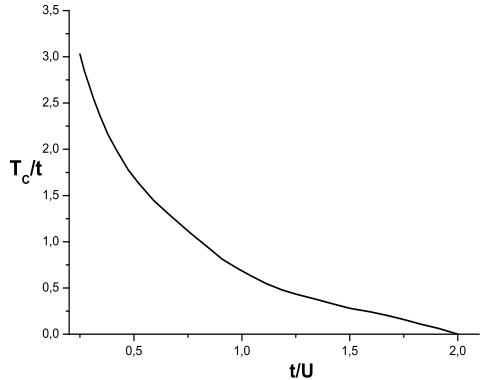,width=8cm,height=6cm}}
\caption{(Colour on-line)\,\,The phase diagram in the plane of temperature $T_C/t$ and control parameter $t/U$. At density of electrons $n=0.4$,  anisotropy parameter $\kappa=-0.83$ and $J'/t=10$ the quantum critical value of the ratio is $t/U=2$.}
\label{QCBFMfig3}
\end{figure}

In the present paper the Hubbard model of itinerant ferromagnetism with transverse anisotropy was studied. Increasing the ratio $t/U$ decreases the effective spin of itinerant electron which in turn decreases the Curie temperature. The equations  (\ref{QCB4d}) and (\ref{QCB5}) show that at fix density of electrons $n=0.4$ the decreasing of the effective spin increases the density of doubly occupied states. The result shows that the quantum critical ground state state is a state with high concentration of doubly occupied states.

This work was partly supported by a Grant-in-Aid DO02-264/18.12.08 from NSF-Bulgaria.

\end{document}